\relax
\documentstyle [prl,aps]{revtex}
\begin {document}
\bibliographystyle {plain}

\title{\bf Comment on ``Nonzero Fermi Level Density of States
for a Disordered d-Wave Superconductor in Two Dimensions'' by
K. Ziegler {\it et al.}}
\maketitle
\sloppy
\par
\bigskip
The paper by Ziegler {\it et al.} 
disputes the results for the density of states of
disordered two dimensional superconductors whose  order parameter 
has $N$ nodes on the Fermi surface, obtained by Nersesyan {\it et
al.}\cite{we}. In this paper the problem was reduced to a problem of
(2 + 1) massless Dirac fermions in a quenched random gauge
potential. Later it was established that such identification is valid
only for $N = 1, 2$, while for $N = 4$ ($d$-wave paring) the situation is
more complicated \cite{we2}, \cite{caux}. The problem of
Dirac fermions in a random gauge field has also been considered by Ludwig {\it et
al.}\cite{ludwig} (the Abelian disorder $N = 1$)  and  by Mudry {\it
et al.}\cite{mudry} (the non-Abelian disorder $N = 2$)  who obtained  the
same 
results as we did, namely that independently of symmetry of the
random gauge field (Abelian or non-Abelian) the density of states at
zero energy vanishes. 

 Though we agree in principle that for $N = 4$ one may have a nonzero
DOS at $E = 0$ (see the discussion in Ref.\cite{caux}), 
we do not think that arguments presented in Ref. 1 have any relevance
to the problem we have studied.   
The starting point for our work was the perturbation theory
result for the density of states at (DOS) at finite energies. 
If the scattering potential is weak, 
the perturbation expansion is essentially independent on whether one
works on a lattice or in the continuous limit.  The expansion contains
logarithmic divergencies $\gamma^n(\ln E)^m$; in three dimensions
leading logs can be summed giving a finite DOS at $E = 0$, but in two
dimensions many extra logarithmic divergencies appear. All these 
logs become of order of one at $E < E_0 = \Delta\exp(-
1/\gamma)$ signifiyng a crossover to strong coupling regime at $|E| << E_0$
\cite{we2}. 
To treat
this  regime we and the other authors suggested 
the  nonperturbative approach. Whether one
agrees with this approach  or not, the  principal fact
of the existence  of extra logarithmic divergencies in the expansion of
DOS at $|E| >> E_0$ cannot be disputed. 

 Now let us look  at the principal formula of the paper \cite{ziegler} -
Eq.(6), which should describe DOS at all energies. Taking $E <<
\Delta$ and calculating the integral yields 
\begin{equation}
N(E) \sim \gamma\ln\left[\Delta^2/(E^2 + \gamma^2)\right] + 
E\tan^{-1}(E/\gamma) \sim  2 \gamma\ln[\Delta/max(E,\gamma)]
\end{equation}
At $E >> \gamma$ the expansion of DOS contains just one logarithm --
in a stricking contrast with the perturbation theory! This is what
should be expected because a Lorentz distribution of disorder, with all
nonvanishing moments divergent,  does not even 
allow to formulate a perturbation expansion. Ziegler {\it et al.}
claim without explicit demonstration  that they can use more general
distributions. We challenge them to reproduce the perturbation theory
result. 

 The discrepancy between Eq.(1) and the
conventional perturbation theory suggests at the very least that models with
Lorentzian and Gaussian disorder belong to different universality
classes (and until  explicit expressions for other distributions
are available there is nothing to discuss). 
However, there may be more than that because 
similar discrepancies exist between the results obtained by this
method for another model of disorder - (2 +
1)-dimensional fermions with random mass \cite{zieg}. This model has
been used to describe the two dimensional Ising model with bond
disorder. The
results obtained within the field theory framework \cite{dot},
\cite{bernard} predict $\ln\ln(T - T_c)$-singularity in the specific
heat in excellent agreement with the computer simulation
made  for  a system of 10$^6$ spins \cite{tal} (the Ziegler's result was
a constant specific heat). The results for the
correlation function averaged over the randomness \cite{lud},
\cite{shenker} also agree with the simulations \cite{tal}. 

 We think that since the authors make the  important claim which goes
well beyond the  argument against one particular result, but is aimed
to debunk  the entire field theoretical approach to disordered
systems, they  should clearly demonstrate that they are dealing
with the same models and  are using  reliable methods. In our opinion,
this has not been achieved in Ref. 1.

\bigskip

A.A.Nersesyan$^1$ and A.M.Tsvelik$^2$\\

$^1$  International Centre for Theoretical Physics, P.O. Box 6586, 34100 Trieste, Italy,\\
and Institute of Physics, Tamarashvili 6, 380077, Tbilisi, Georgia.

$^2$ Department of Physics, University of Oxford, 1 Keble Road, Oxford, OX1 3NP,\\ 
~~~United Kingdom.

\medskip


\end{document}